\documentclass[lettersize,journal]{IEEEtran}

\usepackage{amsmath,amssymb,amsfonts}
\usepackage{pgfplots}
\pgfplotsset{compat=1.18}

\usepackage{graphicx}
\graphicspath{ {figures/} }
\usepackage{subcaption}

\usepackage{algorithm}
\usepackage{algorithmicx}
\usepackage{algpseudocode}

\usepackage{multirow}
\usepackage{booktabs}
\usepackage{tabularx}
\usepackage{array}
\newcolumntype{M}[1]{>{\centering\arraybackslash}m{#1}}
\newcolumntype{Y}{>{\centering\arraybackslash}X}
\usepackage{makecell}

\usepackage{xurl}
\usepackage{hyperref}

\begin{document}

\title{A Comprehensive Study on GDPR-Oriented Analysis of Privacy Policies: Taxonomy, Corpus and GDPR Concept Classifiers}

\author{Peng Tang, Xin Li, Yuxin Chen, Weidong Qiu, Haochen Mei, Allison Holmes, Fenghua Li, and Shujun Li,~\IEEEmembership{Senior Member,~IEEE}%
\thanks{P.~Tang, X.~Li, Y.~Chen, W.~Qiu and H.~Mei are with Shanghai Jiao Tong University, Shanghai, China, 200240. A.~Holmes is with University of Birmingham, UK. S.~Li is with University of Kent, Canterbury, UK. F.~Li is with Institute of Information Engineering, Chinese Academy of Sciences, Beijing, China, 100085.}%
\thanks{Corresponding co-authors: Weidong Qiu (qiuwd@sjtu.edu.cn) and Shujun Li (S.J.Li@kent.ac.uk).}%
}

\maketitle

\begin{abstract}
Machine learning (ML) based classifiers that take a privacy policy as the input and predict relevant concepts are useful in different applications such as (semi-)automated compliance analysis against requirements of a specific data protection law such as the EU GDPR. Although many researchers have studied ML-based privacy policy concept classifiers, we observed multiple research gaps, e.g., the lack of a more complete GDPR taxonomy and the less consideration of hierarchical information in privacy policies. To fill such research gaps, we produced a more complete GDPR-oriented privacy policy concept taxonomy, constructed the first privacy policy corpus with explicitly hierarchical information at three levels, and conducted the most comprehensive performance evaluation study of GDPR concept classifiers for privacy policies, cover many aspects that have not been studied systematically. Our work led to multiple findings and insights, including the usefulness of considering hierarchical contextual features and different hierarchical structures, the observation that a ``one size fits all'' approach may not work, the reduced performance of such classifiers on our newly constructed corpus especially after the first level, and the necessity to split the training and testing sets by documents.
\end{abstract}


\section{Introduction}

\IEEEPARstart{T}{o} provide users with more personalized online services and for other legitimate lawful bases, online services typically collect personal data from their users. For online services provided via the Internet, privacy policies on their websites remain the main type of legal documents for online users to understand how service providers collect their personal data. A typical privacy policy describes different aspects about collection and processing of personal data by an online service, e.g., what personal data are collected, how they are collected, why they are collected, how such data are protected, how such data are stored, and what data are shared with third parties. Although accepting the content of a privacy policy is often made mandatory for starting using an online service, most users tend to omit reading privacy policies because they are often too long to read quickly and too difficult to understand due to the legal and formal wording used~\cite{mcdonald2008cost}. Despite the existence of data protection laws and regulations in many countries, it has been found that service providers' privacy policies often do not fully comply with such laws and regulations, leading to concerns from online users, researchers and privacy advocates~\cite{li2016privacy}. For example, in 2019, the CNIL, the French data protection authority, fined Google 50 million Euros for failing to provide transparent and understandable information in its data consent policy~\cite{Google_GDPR_Violation}.

Among all data protection laws and regulations around the world, the EU (European Union)'s GDPR (General Data Protection Regulation) is one of the most demanding and comprehensive privacy regulations ever enacted. It was passed in 2016, and went into effect in the whole EU and EEA (European Economic Area) on May 25, 2018. After Brexit, the UK decided to keep the GDPR in its local law, known as the UK GDPR, which follows largely the same principles but with some differences on UK-specific matters. This leads to two versions of the GDPR: the EU GDPR and the UK GDPR. In this paper, we will use the loose term ``the GDPR'' to refer to both versions and consider the EU/EEA/UK as the region the GDPR is effective. The GDPR harmonizes data privacy laws across the EU/EEA/UK and is widely regarded as a benchmark for data protection legislation around the world. The GDPR defines a number of principles and requirements for data controllers and data processors to consider in order to be legally compliant. For example, under the GDPR's transparency principle, data subjects have the right to be informed about the collection and processing of their personal data. A common approach to meet this requirement is to provide data subjects with a privacy policy document, which inform them about all important information they have a right to know according to the GDPR. The GDPR requires each country to have a national authority to take care of the enforcement of the GDPR, and such bodies often release guidelines to data controllers and data processors on how to be GDPR compliant. For instance, the Information Commissioner's Office (ICO)~\cite{ICO_website}, the UK's national data protection authority in charge of the enforcement of the GDPR\footnote{The EU GDPR in the UK before Brexit and the UK GDPR afterwards.}, has provided a general guide on the GDPR~\cite{ICO_GDPR_guide} and also a template for data controllers to use as a reference of GDPR-compliance privacy policies~\cite{ICO_privacy_template}. In order to meet their legal obligation in terms of the GDPR, many organizations updated their privacy policy to be GDPR compliant, e.g., the New York Times updated its privacy policy on May 24, 2018, the day before the GDPR went into effect, to include provisions on international data transfers.

The importance of privacy policies for both data subjects and data controllers means that GDPR-oriented analyses of such legal documents can be very useful. It can help enhance data subjects' awareness on important data protection issues, and also help data controllers refine their privacy policy to be more legally compliant to the GDPR. Such an analysis can obviously be done qualitatively by experienced GDPR experts, but more automated analysis is preferred in many applications to help reduce the time and costs spent. One frequently studied privacy policy analysis task is concept classification, which involves classifying a given segment of a privacy policy into one or more concepts (see Section~\ref{subsec:related_work:PPCC} for a brief literature review).
We observed multiple research gaps in the literature on privacy policy concept classification, including the lack of a more complete GDPR taxonomy and a privacy policy corpus that include more semantic levels and hierarchical information, the lack of consideration of hierarchical information in privacy policies for developing concept classifiers, the lack of studies comparing classifiers with different  hierarchical structures, and the inconsistent use of sample splitting methods for generating training and testing sets.

This paper reports our work which helps fill the above research gaps with the following \textbf{key contributions} (ordered following the logical sequence of the contributions).

\emph{1) We produced an extended taxonomy for facilitating GDPR-oriented analysis of privacy policies.} This taxonomy is based on a smaller GDPR taxonomy proposed by Torre et al.~\cite{torre2020ai}, our own knowledge on the GDPR as cyber law experts, two important documents from the ICO and the IAPP (International Association of Privacy Professionals), and the work of the W3C Data Privacy Vocabularies and Controls CG (DPVCG). The end result is the most comprehensive GDPR privacy policy taxonomy reported in the literature so far.

\emph{2) We constructed GoPPC-150, the first fully hierarchically encoded GDPR-oriented privacy policy corpus, and a new privacy policy corpus construction framework we used to construct GoPPC-150.} GoPPC-150 includes 150 privacy policies collected from Alexa.com top websites. It includes expert-annotated GDPR concept labels in a hierarchical manner following a newly extended GDPR privacy policy taxonomy (see below), and is structured at the document-level. The inclusion of explicitly encoded hierarchical information in GoPPC-150 allows more context-aware development of GDPR concept classifiers and other relevant tools. The framework developed has a high level of automation to minimize human intervention and can be used to further extend GoPPC-150 and to construct other similar corpora.

\emph{3) We conducted the most comprehensive performance evaluation study on GDPR concept classifiers (to the best of our knowledge).} Our study goes beyond past research by considering the following aspects: i) two hierarchical architectures of such classifiers -- local classifier per node (LCN) and local classifier per parent node (LCPN), not just LCPN in past work; ii) multiple sets of traditional features and contextual features reflecting the hierarchical nature of a privacy policy; iii) different machine learning models, and iv) two different sample splitting methods (document- and segment-level). In addition to being the most comprehensive performance evaluation study, we are the first who considered the hierarchical nature of such  classifiers and relevant features and investigated the impact of different sample splitting methods on the classification performance.

\emph{4) Our comprehensive performance evaluation study led to a range of new findings and insight.} These include the usefulness of considering hierarchical contextual features and different hierarchical structures, the observation that a ``one size fits all'' approach may not work, the reduced performance of such classifiers on our newly constructed corpus especially after the first level, and the necessity to use the document-level sample splitting method for producing the training and testing sets.

\begin{figure}[!t]
\centering
\includegraphics[width=\linewidth]{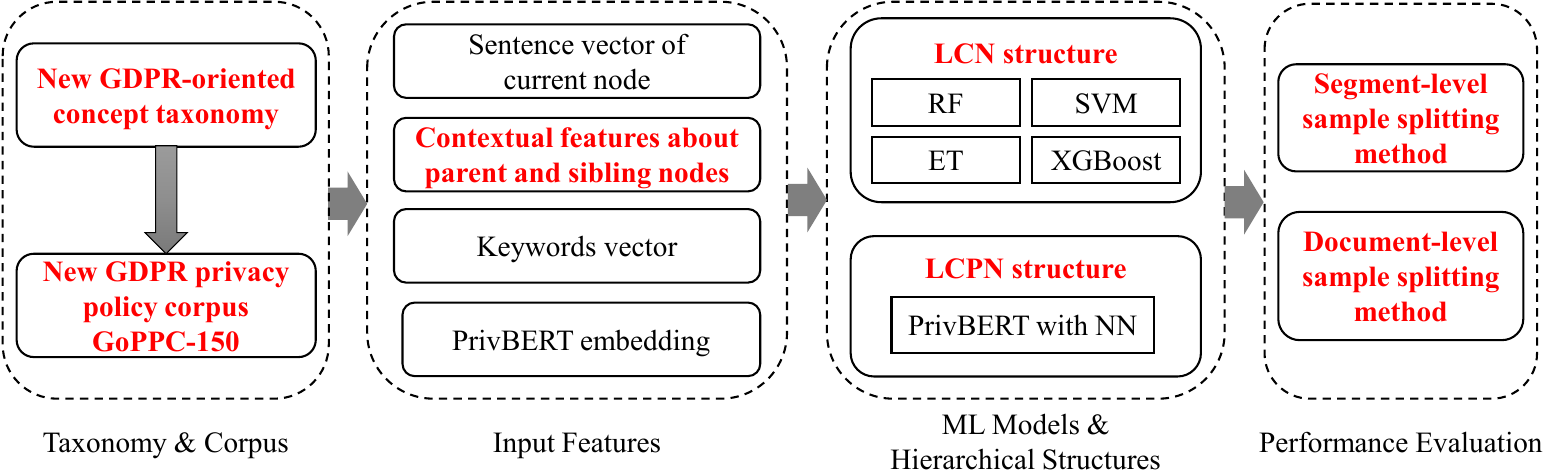}
\caption{A diagram showing different parts of our work reported in this paper.}
\label{fig:comprehensive_study_concept_classifiers}
\end{figure}

Figure~\ref{fig:comprehensive_study_concept_classifiers} visually shows how different parts of our work and key contributions are organized, with the components highlighted in red indicating the new aspects investigated in our work. To help other researchers reproduce our work reported in this paper, all our source code, data used and newly produced, and all results of our experiments have been uploaded to a GitHub repository available at \url{https://github.com/tp-sh/GDPR_privacy_policies}. We will refer to this GitHub repository frequently in the rest of the paper, with a direct URL pointing to the corresponding part of the repository when necessary.

The rest of this paper is structured as follows. Section~\ref{sec:related_work} discusses related work.
Section~\ref{sec:taxonomy_corpus} presents the new GDPR-oriented taxonomy and the new privacy policy corpus we developed, detailing the methodology used in their development. Section~\ref{sec:classifiers_evaluation} presents our comprehensive performance evaluation of GDPR concept classifiers using our new privacy policy corpus, covering experimental settings and results. Section~\ref{sec:further_discussions} provides further discussions of our work and also indicates some future research directions. The last section concludes the paper.

\section{Related Work}
\label{sec:related_work}

In this section, we introduce related work in four closely related areas: (1) hierarchical multi-label text classification, (2) privacy policy corpora, (3) privacy policy concept classifiers, and (4) analyses of privacy policies. How our work is compared with some selected related work can be found in Table~\ref{table:selected_related_work}.

\newcommand\ssdagger{\textsuperscript{\textdagger}}
\begin{table*}[!htb]
\centering
\footnotesize
\caption{Our work compared with related work (* =  new public corpus; \ssdagger\ = GDPR-specific; texts in bold face = key contributions of our work). The tuples in the ``Taxonomy'' column indicate the numbers of levels and nodes of the taxonomies, e.g., (3, 96) refers to a taxonomy with 96 nodes at 3 levels. The tuples in the ``Corpus/Corpora'' column indicate the number of privacy policies, the number of annotation levels and if the corpus includes hierarchical information, e.g., (150, 3, Yes) means that there are 150 privacy policies with annotations at 3 levels, and hierarchical information is included in the annotations. Note that when counting the number of levels, the root node is not counted.}
\label{table:selected_related_work}
\begin{tabularx}{\linewidth}{cccccY}
\toprule
Related Work & Taxonomy & Features & Corpus/Corpora & Classifier(s) & Sample Splitting Method(s)\\
\midrule
OPP-115~\cite{Wilson2016WPPcorpus} & (2, 46) & Paragraph2Vec & OPP-115* (115, 1, No) & Multiple; LCPN & Document-Level\\
Polisis~\cite{Harkous2018Polisis} & No & fastText & OPP-115 & CNN; LCPN & Document-Level\\
Mustapha et al.'s~\cite{Mustapha2020PPC-XLNet} & No & XLNet & OPP-115 & XLNet; LCPN & Segment-Level\\
Mousavi Nejad et al.'s~\cite{Nejad2020baselinePPC} & No & fastText+BERT & OPP-115 & CNN+BERT; LCPN & Segment-Level\\
PrivBERT~\cite{Srinath2020PrivaSeerPrivBERT} & No & RoBERTa & OPP-115 & CNN; LCPN & Segment-Level\\
Torre et al.'s~\cite{Amaral2022} & (3, 69)\ssdagger & GloVe & Private\ssdagger & SVM\ssdagger; LCN & Document-Level\\
Rahat et al.'s~\cite{rahat2021automated} & (1, 18)\ssdagger & fastText & Private & CNN\ssdagger; LCPN & Document-Level\\
\midrule
\textbf{Our Work} & (3, \textbf{96})\ssdagger & \textbf{Multiple}; \textbf{Context} & \textbf{GoPPC-150*\ssdagger\ (150, 3, Yes)} & \textbf{Multiple}; LCPN \textbf{\&} LCN\ssdagger & Document-Level \textbf{\&} Segment-Level\\
\bottomrule
\end{tabularx}
\end{table*}

\subsection{Hierarchical Multi-Label Text Classification (HMTC)}
\label{subsec:related_work:HMTC}

Privacy policy is a hierarchical text with one or more labels for each sentence / paragraph, so the concept classification task of a given privacy policy can be regarded as an HMTC task. According to Silla et al.~\cite{silla2011hierarchical}, there are three main approaches to HMTC -- flat (learning about leaf nodes only so completely ignoring the hierarchical structure), local (learning about each node or sets of nodes separately) and global (learning all nodes at the same time) classifiers. Flat classifiers are naive and global ones are usually too complicated and require sufficient training data for all nodes. Therefore, local classifiers have received the most attention from researchers, and multiple types have been developed, including LCN (a local classifier per node), LCPN (a local classifier per parent node) as well as LCL (a local classifier per level) classifiers~\cite{costa2007comparing, fagni2007selection, secker2010hierarchical}.

When it comes to privacy policy concept classification, most researchers also paid more attention to local classifiers~\cite{Harkous2018Polisis, Mustapha2020PPC-XLNet, torre2020ai, rahat2021automated, Srinath2020PrivaSeerPrivBERT}, particularly LCPN classifiers based on CNN. The ``Classifier(s)'' column of Table~\ref{table:selected_related_work} shows the type of classifiers studied in some selected related work. As can be seen, LCN classifiers have been much less studied and we were unaware of any studies comparing performances of LCPN and LCN classifiers. LCL classifiers are usually not appropriate for privacy policy concept classification since concepts belonging to different branches are often too different to be processed by a single classifier. This is why we have not seen any related work using LCL classifiers.

\subsection{Privacy Policy Corpora}

In 2016, Wilson et al.~\cite{Wilson2016WPPcorpus} created a public privacy policy corpus called ``Online Privacy Policy'' (OPP-115) by hiring three legal experts as annotators and it has been used by many other researchers. However, Sarne et al.~\cite{sarne2019unsupervised} used unsupervised ML to model topics in privacy policies and found a mismatch between the topics in privacy policies they analyzed and topics covered in OPP-115. Sathyendra et al.~\cite{sathyendra2017identifying} constructed a more fine-grained corpus based on OPP-115 with semi-automated annotation, focusing on ``opt-out'' in privacy policies. Leblanc and Liu~\cite{lebanoff2018automatic} built a corpus of fuzzy words and sentences in privacy policies. Zimmeck et al.~\cite{zimmeck2019maps} created the App-350 Privacy Policy Corpus with the goal of checking compliance of behaviors and privacy policies of mobile apps. They selected 350 privacy policies of the most popular apps on Google Play and hired legal experts as annotators. Robaldo et al.~\cite{robaldo-etal-2020-dapreco} used I/O logic formula to code and model various legal documents, expressed legal statements in a logical language, explained various clauses, and demonstrated their work with the GDPR. Muller et al.~\cite{muller2019gdpr} introduced a privacy policy dataset containing over 18,300 sentences, tagged according to five core privacy policy requirements of the GDPR. Srinath et al.~\cite{Srinath2020PrivaSeerPrivBERT} created PrivaSeer, an automatically constructed corpus containing privacy policies of over 1 million English websites, and studied the composition of the corpus, showing readability tests, document similarity, keyword extraction results, and exploring the corpus through topic modeling. Kuznetsov et al.~\cite{Kuznetsov2022lotdataset} used a technique for identifying URLs of privacy policies of IoT devices to construct a new corpus with 592 privacy policies in 2022. Zhan et al.~\cite{Zhan2024VPVet} constructed VRPP, the first large-scale dataset of privacy policies for virtual reality applications, comprising 11,923 distinct VR apps from 10 mainstream platforms.

\subsection{Privacy Policy Concept Classifiers}
\label{subsec:related_work:PPCC}

Harkous et al.~\cite{Harkous2018Polisis} proposed Polisis, a comprehensive framework for enabling multi-dimensional privacy policy analysis. Polisis provides an online service for privacy policies analysis, using a combination of NLP and deep learning (DL) techniques to extract fragments from privacy policies, each containing a set of labels describing data processing behaviors. Tesfay et al.~\cite{tesfay2018privacyguide} proposed an ML-based approach to classify multiple categories of privacy policy content using pre-defined keywords. Lippi et al.~\cite{lippi2019claudette} provided 33 metadata types for privacy policies in terms of the GDPR compliance, and provided automatic support for ambiguity detection of privacy policies based on manual rules and machine learning (ML) based methods. A more comprehensive piece of work was done by Torre et al.~\cite{torre2020ai}, who proposed 55 metadata types, covering all types identified by Lippi et al.~\cite{lippi2019claudette}, and studied automated detection of privacy policy integrity using an advanced combination of natural language processing (NLP) and ML based on the 55 metadata types. Their more recent work~\cite{Amaral2022} further improved their privacy policy completeness checking framework with the same GDPR taxonomy and classifiers. Mousavi Nejad et al.~\cite{Nejad2020baselinePPC} studied the performance of multiple machine learning techniques (CNN and two BERT-based transformers) on classifying privacy privacy concepts in the OPP-115 corpus, and reported the best performance figures of 77–85\% (micro-avg) and 76–79\% (macro-avg). Mustapha et al.~\cite{Mustapha2020PPC-XLNet} proposed to use a pre-trained XLNet and a dense layer for classification of concepts in OPP-115, and showed the classifier can outperform the best models reported in~\cite{Nejad2020baselinePPC} by 1-3\% for both macro and micro average F1-scores. Srinath et al.~\cite{Srinath2020PrivaSeerPrivBERT} constructed a new privacy policy corpus called PrivSeer, built a pre-trained language model called PrivBERT based on PrivSeer, and then developed a number of concept classifiers for privacy policy analysis based on PrivBERT. As a whole, the best-performing classifiers are PrivBERT-based ones and those reported in~\cite{Amaral2022}, achieving an F1 score above 0.8 for most concept classifiers, and even above 0.9 for some. Xie et al.~\cite{Xie2025LLMPrivacy} constructed a collection of privacy policy clauses grounded in multiple privacy regulations, leveraging self-annotated excerpts from both open-source datasets and real-world policies, and applied zero-shot prompting with large language models for clause classification.

\subsection{Analyses of Privacy Policies}

In addition to privacy policy concept classification, there are many other past studies focusing on automated or semi-automated analyses of privacy policies, some of which are GDPR-oriented. Many of such studies are based on machine learning based concept classifiers, as part of a larger automated system.

When it comes to the GDPR-oriented analyses, Tom et al.~\cite{tom2018conceptual} presented a preliminary GDPR model that aims to provide a simple and visual overview for human operators to achieve a better understanding of the relationships between different concepts in the GDPR. It also describes a method of using their proposed model as a tool to develop privacy policies and illustrates how to extract compliance rules. Palmirani and Governatori~\cite{palmirani2018modelling} proposed a proof of concept applicable to the GDPR domain with the aim of detecting or preventing violations of privacy enforcement norms. Bhatia et al.~\cite{bhatia2019identifying} identified incompleteness of privacy policies by representing data practice descriptions as semantic frames. The approach was a grounded analysis to discover which semantic roles corresponding to a data action are needed to construct complete data practice descriptions. Mousavi et al.~\cite{nejad2018knight} proposed a tool called KnIGHT, whose innovation lies in the use of semantic similarity between words to associate sentences in a privacy policy with relevant paragraphs in the legal text of the GDPR. Torre et al.~\cite{torre2019using} proposed a model-based solution for GDPR-compliance analysis, using Unified Modeling Language (UML) and Object Constraint Language (OCL) to build a UML representation of the GDPR. Hamdani et al.~\cite{Hamdani2021GDPRcompliance} conceptualized a framework to implement document-central compliance checking methods in the data supply chain and developed concrete methods to automatically check GDPR-compliance of privacy policies. Xiang et al.~\cite{Xiang2023PolicyChecker} developed PolicyChecker, a rule-based framework that encodes the mandatory and context-dependent requirements from GDPR Articles 13 and 14 into logical chains to automatically assess the completeness of mobile app privacy policies, evaluated on 205,973 Android apps from the UK Google Play Store.

Some other work focused on the commonalities of privacy policies. In order to increase transparency of privacy policies, Zimmeck et al.~\cite{Privee} proposed Privee, a software architecture for analyzing essential policy terms based on crowd sourcing and automatic classification techniques. They implemented Privee as a proof-of-concept web browser extension that retrieves policy analysis results from an online privacy policy repository or, if no such results are available, performs automatic classifications. Targeting privacy policies of websites and mobile apps, Caramujo et al.~\cite{caramujo2019rsl} proposed a domain-specific language and a model transformation approach for specifying privacy policy models. Pullonen et al.~\cite{pullonen2019privacy} proposed a multi-level model as an extension of the business process model and representation to support visualization, analysis, and communication of the privacy policy characteristics of business processes. Ayala-Rivera and Pasquale~\cite{ayala2018grace} proposed a model-based approach to help online services understand their data protection obligations under the GDPR. Zimmeck et al.~\cite{2021PrivacyFlash} designed and implemented PrivacyFlash Pro, an automated privacy policy generator for iOS apps that leverages static analysis, which identifies code signatures composed of Plist permission strings, framework imports, class instantiations, authorization methods and other evidence that are mapped to privacy practices expressed in privacy policies. Zhou et al.~\cite{Zhou2023POLICYCOMP} proposed POLICYCOMP, an automated framework that identifies overbroad personal data collection practices in mobile app privacy policies by comparing them against functionally similar counterpart apps, revealing that nearly half of such practices may exceed actual service needs. Wang et al.~\cite{wang2022privguard} proposed PrivGuard, a novel system design that reduces human participation required and improves the productivity of the compliance process, which is mainly comprised of two components: (1) PrivAnalyzer, a static analyzer based on abstract interpretation for partly enforcing privacy regulations, and (2) a set of components providing strong security protection on the data throughout its life cycle.

Some work has a focus on helping users to better comprehend privacy policies, especially via visual and graph-based mechanisms. Cui et al.~\cite{cui2023poligraph} proposed PoliGraph, automated analysis of the information disclosed in a policy into a knowledge graph based on NLP technology. Although they also combine the context of the current text for classification, they only focus on the data collection part of the privacy policy, and do not consider the structural information of the privacy policy. To enhance user understanding of data handling implications, Zimmermann et al.~\cite{Zimmermann2025VisualPP} conducted a between-subjects study, evaluating the effectiveness of visual metaphors and dynamic feedback in supporting informed privacy decisions. Pan et al.~\cite{Pan2024ANH} proposed SeePrivacy, a multimodal framework that combines GUI visual analysis and large language models to extract context-relevant segments from privacy policies in mobile applications, aiming to enhance user interaction with and understanding of those policies. Adhikari et al.~\cite{Adhikari2025PolicyPulsePS} introduced PolicyPulse, an NLP pipeline that focuses on extracting semantic roles of different data entities from a privacy policy to enhance users' comprehension of privacy policies. The role analysis is achieved by first classifying different semantic frames into relevant data practice categories, and then relevant frames are further processed to extract semantic roles. The frame categorization step is very similar to the concept classification discussed in the previous subsection, but PolicyPulse only involves 5 selected categories from OPP-115 that are more useful for supporting the semantic role extraction. This work therefore is not directly about concept classification in privacy policies. Yang et al.~\cite{Yang2025PPAnalysisLLM} proposed a framework that uses large language models and knowledge distillation to automatically extract structured privacy attributes from policies and represent them as queryable graphs.

\section{Extended GDPR Taxonomy and New Corpus}
\label{sec:taxonomy_corpus}

In this section, we give details of our work on the extended GDPR taxonomy, the new corpus GoPPC-150 and the framework we developed to construct the corpus.

\subsection{Extended GDPR Taxonomy}
\label{subsec:GDPR_taxonomy}

In 2020, Torre et al.~\cite{torre2020ai} proposed a GDPR-oriented, three-level conceptual model of privacy policy metadata by cooperating with legal experts. Although being professionally constructed, we noticed that Torre et al.'s model still has a number of issues, including some missed concepts and structural issues of some nodes. So we decided to refine it further by making a range of changes, leading to an extended taxonomy covering a more comprehensive set of GDPR-related concepts relevant for privacy policies. The extension was based on the GDPR-oriented privacy policy template provided by the ICO, the privacy policy mapping chart provided by the IAPP~\cite{IAPP2021Privacy_Law_Mapping_Chart}, the work of W3C Data Privacy Vocabularies and Controls CG (DPVCG)~\cite{DPVCG}, the annotation scheme of the OPP-115 corpus~\cite{Wilson2016WPPcorpus}, and our own expertise\footnote{Two co-authors of the paper have been actively teaching and researching the GDPR. One of them is a data protection and privacy law expert, and the other is a computer scientist with substantial research experience on interdisciplinary topics including GDPR-related matters.}. Our extended taxonomy includes 96 nodes, 39.1\% more than Torre et al.'s (69 nodes), including some important nodes missing from the latter, e.g., `DATA SHARING' and the sub-nodes of `PD STORAGE DETAILS'.

The main changes we made to Torre et al.'s original model and the reasons of such changes are summarized below.

\textbf{Change 1:} Swapped `AUTO DECISION MAKING' and `DATA SUBJECT RIGHT.COMPLAINT'.

\textbf{Reason:} The ICO guide has ``Rights related to auto decision making including profiling'' under the ``Individual rights'' category, and ``complaints'' is covered in a dedicated section in the ICO's GDPR-oriented privacy policy template.

\textbf{Change 2:} A new `CONDITION' node was added, which was combined with the original `RECIPIENTS' under the new first-level node `DATA SHARING'.

\textbf{Reason:} When describing data sharing, privacy policies often include conditions for sharing.

\textbf{Change 3:} Changed `PD TIME STORED' node to `TIME', and merged the newly added `LOCATION' node and `DISPOSAL METHOD' node into the new first-level node `PD STORAGE DETAILS'.

\textbf{Reason:} These changes reflect the recommended content under ``data storage'' in the ICO template better.

\textbf{Change 4:} Split the `DIRECT' sub-node of `PD ORIGIN' into `DIRECT ACTIVE' and `DIRECT PASSIVE', and added third-level node `COOKIE' under `INDIRECT' to the scope of `DIRECT PASSIVE'.

\textbf{Reason:} When describing data collection, privacy policies often mention data provided by users (`DIRECTIVE ACTIVE'), data automatically collected by service providers (`DIRECTIVE PASSIVE'), and data from third-party sources (`INDIRECT'). Torre et al.'s~\cite{torre2020ai} taxonomy included just `DIRECT' and `INDIRECT' nodes, which cannot accurately cover data automatically collected by service providers (`DIRECTIVE PASSIVE'). Cookies as a type of data automatically collected would now better be put under `DIRECTIVE PASSIVE'.

\textbf{Change 5:} Added the `INFORMATION' node to the `DATA SUBJECT RIGHT'.

\textbf{Reason:} Articles~13 and 14 of the GDPR declare that a range of specific information ``shall'' be provided by data controllers.

\textbf{Change 6:} Some changes were made after comparing our taxonomy with the IAPP~\cite{IAPP2021Privacy_Law_Mapping_Chart}. For example, processing of personal data by online services must satisfy some principles. Some of them are internal and unnecessary to be declared to users, but the others are external and need to be declared in privacy policies. We added a new first-level node called `DP PRINCIPLE', containing `PURPOSE LIMITATION' and `DATA MINIMIZATION'.

\textbf{Change 7:} For concepts that go beyond the GDPR, we added two new first-level nodes to the taxonomy: `NON-GDPR' and `OTHERS'. `NON-GDPR' refers to concepts related to data protection laws in other countries or regions, such as the California Consumer Privacy Act (CCPA) in the US~\cite{CCPA}, Brazil's General Personal Data Protection Law (Lei Geral de Proteção de Dados Pessoais, LGPD)~\cite{LGPD}, and the Personal Information Protection Law of China~\cite{ChinaPIPL}. `OTHERS' covers all situations that cannot be labeled.

In addition to the above changes, we also compared our taxonomy with the work of W3C Data Privacy Vocabularies and Controls CG (DPVCG)~\cite{DPVCG}, and confirmed that all key concepts in the latter are covered in our extended taxonomy.

\begin{figure*}[!htb]
\centering
\includegraphics[width=0.8\linewidth]{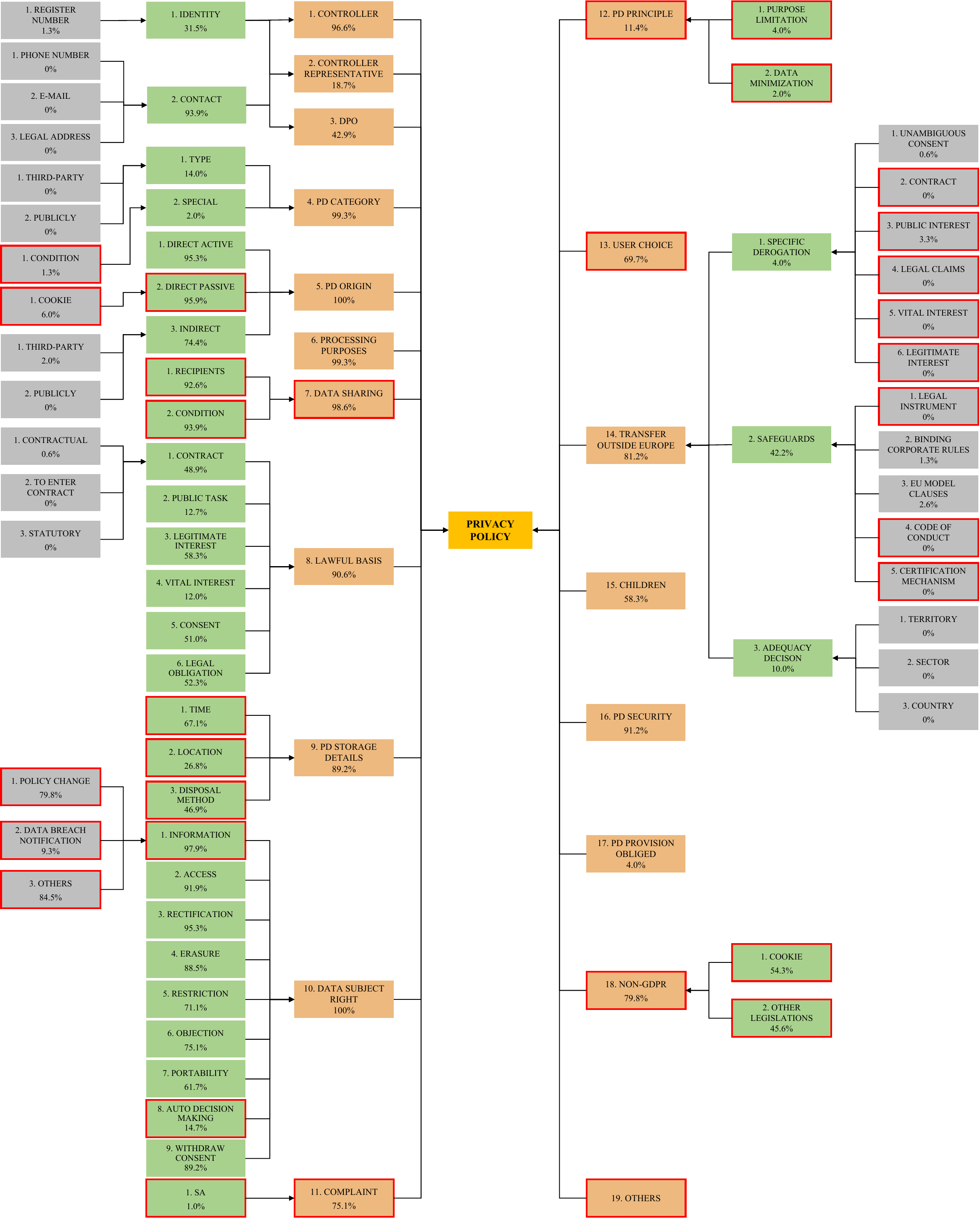}
\caption{The proposed GDPR-oriented privacy policy taxonomy (red boxes = newly added concepts, percentages = coverage rates of different GDPR concept in the our new GoPPC-150 corpus).}
\label{fig:GDPR_taxonomy}
\end{figure*}

According to the changes described above, we obtained a more comprehensive GDPR taxonomy, which lays the foundation of other work reported in this paper. As shown in Figure~\ref{fig:GDPR_taxonomy}, our GDPR taxonomy is a three-level tree covering more important core GDPR concepts relevant for privacy policies. In order to enhance the readability of the visual presentation, different levels are colored differently and nodes we added or modified are highlighted inside red boxes. And the GDPR concepts with underlines are those covered in the ICO template. The percentage under the nodes are the coverage rates of different GDPR concept in the our new GoPPC-150 corpus. The results are as follows.

Out of the 19 first-level concepts, `PD ORIGIN' and `DATA SUBJECT RIGHT' are covered by all privacy policies, followed by `PD CATEGORY' and `PROCESSING PURPOSES' (all but one), indicating that these four aspects of the GDPR have been taken seriously by most websites covered in our corpus. On the other hand, many other first-level GDPR concepts have a much lower coverage rate, e.g., `PD PROVISION OBLIGED' covered by only six (4.0\%) privacy policies, despite that this concept is required to appear in privacy policies according to Article~13.2(e) of the GDPR (where the mandatory wording ``shall'' is used).

For the second-level concepts, `DATA SUBJECT RIGHT.INFORMATION', `PD ORIGIN.DIRECT PASSIVE' and `PD ORIGIN.DIRECT ACTIVE' have the highest coverage, mentioned in 146 (97.9\%), 143 (95.9\%) and 142 (95.3\%) privacy policies, respectively. On the low coverage end, `PD CATEGORY.SPECIAL', `COMPLAINT.SA' and `PD PRINCIPLE.DATA MINIMIZATION' have the lowest coverage rate of 2.0\%, covered in only 3 privacy policies.

Among the third-level concepts, only `DATA SUBJECT RIGHT.IN\-FORMATION.POLICY CHANGE' has a relatively high coverage, mentioned in 119 (79.8\%) privacy policies, other third-level concepts are rarely mentioned. A possible explanation is that online services may consider such low-level details in privacy policies unnecessary.

\subsection{The New Corpus GoPPC-150}
\label{sec:corpus_construction}

Privacy policies of different websites and online services can vary significantly in many ways, e.g., in terms of their length, complexity, presentation format, and legal compliance requirements. Generally speaking, privacy policies of large companies are written and maintained by a dedicated legal team, which are likely to be more comprehensive and professionally written. For the new corpus we constructed, our aim is to cover typical privacy policies used by large companies, so we decided to use top sites returned by Alexa.com, a widely used website ranking online service for research purposes\footnote{This service was discontinued by in May 2022.}, for collecting privacy policy samples. The full list of the 150 websites is provided in the supplementary data file available at \url{https://github.com/tp-sh/GDPR_privacy_policies/blob/main/data/Web_list.csv}. Different from former corpora, our new corpus is the first and the only one maintaining the hierarchical structure of original privacy policies with annotations covering three levels. The corpus contains manual annotations of 8,356 fine-grained paragraphs and 5347 titles in the 150 privacy policies.

\begin{figure*}[!t]
\centering
\includegraphics[width=0.85\linewidth]{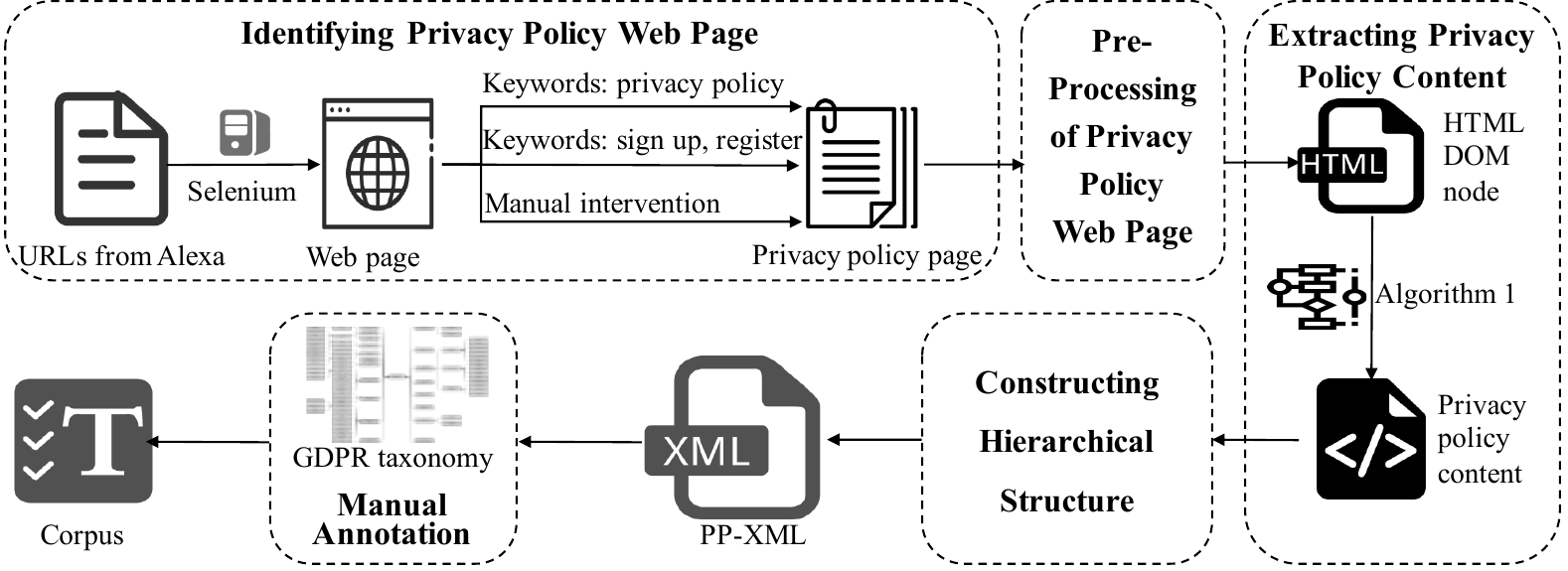}
\caption{The architecture of the framework for constructing GoPPC-150}
\label{fig:architecture_GoHPPC}
\end{figure*}

We developed the framework shown in Figure~\ref{fig:architecture_GoHPPC} to construct our corpus GoPPC-150, which can also be used to further extend GoPPC-150 and to facilitate construction of other similar corpora. The purpose of the framework is to support more automated identification of privacy policy web pages from a pre-defined list of URLs, extraction of the privacy policy's relevant content, and converting the HTML elements in the privacy policy web page into a hierarchical structure following a well-defined XML schema we call PP-XML (privacy policy XML). The framework helps reduce human efforts greatly, and can be adapted to enhance automation of privacy policy analysis.

\subsubsection{Identifying Privacy Policy Web Page}
\label{sec:privacy_policy_extraction}

We observed that some websites change the content or even URL of their privacy policy according to the country or region a visitor comes from. In order to ensure that the collected privacy policies for our corpus are more relevant for the GDPR, we used a proxy server called Amazon EC2 located in London to simulate visits of a user from the EU/EEA/UK to the selected candidate websites. We developed a semi-automated process for identifying and downloading privacy policies from a list of pre-defined websites. The tool was implemented based on the web browser automation engine Selenium~\cite{Selenium} and it works following the steps described below.

\textbf{Step 1}: The tool visits each website and searches (case-insensitively) for a \verb|<a>| element whose content includes one of the following pre-defined keywords representing a privacy policy: `privacy policy', `privacy notice', and `privacy terms'. If only one link is found, the tool clicks the link to visit the privacy policy web page and goes to Step 4; otherwise, it goes to Step 2.

\textbf{Step 2}: The tool tries to identify the privacy policy web page via the user registration page, which normally includes a link to the privacy policy. To this end, it searches for the following keywords case-insensitively: `create account', `register', `sign up', `sign-up'. If any of the keywords is found, it clicks the link and goes to the user registration page. Then, the tool applies Step 1 to the current page to find and visit the privacy policy page, after which it goes to Step 4. If no privacy policy link is found, the tool goes to Step 3.

\textbf{Step 3}: The tool seeks human intervention to identify the link of the privacy policy.

\textbf{Step 4}: The tool saves the identified privacy policy page as a local HTML file with all elements included, by calling a third-party web browser extension SingleFile~\cite{SingleFile@GoogleChromeWebStore}.

\subsubsection{Pre-Processing of Privacy Policy Web Page}
\label{subsubsec:pre-processing-PP-web-page}

After identifying a privacy policy web page, we need to pre-process the web page to remove three types of DOM elements: 1) multimedia-related elements that are not useful for analyzing textual content of the privacy policy contained in the web page, e.g., \verb|<picture>|, \verb|<img>|, \verb|<video>|, and \verb|<audio>|; 2) elements that embed an non-textual object or an external web page, e.g.,  \verb|<applet>|, \verb|<embed>|, \verb|<object>|, and \verb|<iframe>|; 3) elements whose are not semantically related to or can appear as part of the main body of a web page, e.g., \verb|<footer>| and \verb|<nav>|. Note that some of such elements are still useful for understanding the design and layout of a privacy policy web page, which is out of the scope of this study and will be our future work. The full list of HTML elements removed is shown in Appendix~\ref{appendix:HTML_elements_removed}.

\subsubsection{Extracting Privacy Policy Content}
\label{subsubsec:extracting_PP_content}

After pre-processing a privacy policy web page, our next step is to extract the relevant content of the privacy policy. Here, the term ``relevant content'' refers to one or more HTML elements that contain the actual content of a privacy policy. Inspecting all the 150 privacy policy web pages we collected, we observed that the relevant content is always under a single HTML element with multiple child elements each corresponding to a different part of the privacy policy (e.g., a \verb|<div>| element including $n$ child \verb|<div>| elements). In this way, we need to identify the single HTML element that contains the content of the privacy policy, which we call the PP element.

We observed that the immediate child elements under the PP element are more similar to each other as a whole in terms of the text length within each element, compared with non-PP elements at the same level. Here, the term ``text length'' is defined as the number of characters for human readers (not the characters in the HTML code), reflecting the amount of text a human reader is expected to read. Based on the above observations, we developed an algorithm shown in Algorithm~\ref{algo:MPN_webpage}. It measures the likelihood of a candidate element being the PP element using a simple metric we call children similarity score (CSS) -- the standard deviation of all its children's text lengths. Since the CSS is a relative concept, Algorithm~\ref{algo:MPN_webpage} compares the CSS value of the current candidate element with the average CSS value of all non-PP elements observed so far, and if the ratio is below a threshold $r_h$, the candidate element is considered the PP element.

\begin{algorithm}[!htb]
\caption{Given an HTML DOM node (normally \texttt{<body>}), return an HTML element that is more likely the lowest node containing the privacy policy's relevant content fully}
\label{algo:MPN_webpage}
\begin{algorithmic}[1] 
\Function{Extraction}{node}
    \State $\mathbb{R} \gets \varnothing$
    \State $r \gets \infty$
    \State PP\_node $\gets$ node
    \While{True}
         \If{PP\_node does not have any child node}
            \State return PP\_node
         \EndIf
         \State $s \gets$ the CSS value of PP\_node
        \If{$\#(\mathbb{R})>0$}
            \State $r \gets s/\text{average}(\mathbb{R})$
        \EndIf
        \If{$r < r_h$}
            \State \Return{PP\_node}
        \Else
            \State $\mathbb{R} \gets \mathbb{R}\bigcup\{s\}$
            \State PP\_node $\gets$ the child node in $\mathbb{C}$ with the longest human-readable text
        \EndIf
    \EndWhile
 \EndFunction   
\end{algorithmic}
\end{algorithm}

We carried out experiments in 150 samples collected in Section~\ref{sec:privacy_policy_extraction} to test the performance of Algorithm~\ref{algo:MPN_webpage}. In order to determine the optimal value of the parameter $r_h$, we used 50 randomly selected samples as the training set to find its range that allowed us to achieve a zero identification error for the 50 samples. The range obtained is [0.5,0.6], and we assigned $r_h$ to be the midpoint 0.55. Applying Algorithm~\ref{algo:MPN_webpage} with $r_h=0.55$ to the other 100 remaining policies as the testing set and achieved a perfect accuracy of 100\%.

Despite the high accuracy of Algorithm~\ref{algo:MPN_webpage}, it may still fail when processing an unseen privacy policy (e.g., a privacy policy web page does not have a single HTML element that contains the relevant content), so results of this step should be checked and fixed manually when necessary.

\subsubsection{Constructing Hierarchical Structure}
\label{subsubsec:constructing_hierarchy}

After extracting the relevant content of a privacy policy web page, we convert the HTML-based DOM tree of the page content into a new hierarchical structure representing the semantic content of the privacy policy, which is stored as an XML file following a specific XML schema we call PP-XML. The whole privacy policy content has a single root node \verb|<policy>|, which includes a number of semantic segments of the privacy policy, each encoded as a \verb|<segment>| element. The \verb|<segment>| elements can be nested to allow a hierarchical structure for privacy policy's content. For each \verb|<segment>| element, there is always one \verb|<title>| element, which represents the semantic heading of the corresponding segment's content. Each \verb|<segment>| element should have one or more \verb|<paragraph>| elements, representing the content of the corresponding segment. In addition to \verb|<segment>| elements, the \verb|<policy>| element can also include a number of standalone \verb|<paragraph>| elements, e.g., for a number of leading paragraphs before the first \verb|<segment>| element. A \verb|<paragraph>| element can also include one or more \verb|<list>| elements as its children, and a \verb|<list>| element contains at least one \verb|<item>| element as its children. An \verb|<item>| element may include a \verb|<list>| element, allowing nested lists in a \verb|<paragraph>| element.

Note that there has been some related work on automatic segmentation of HTML documents, e.g., the ASDUS proposed in~\cite{gopinath2018supervised}. However, most such work focused on identifying top-level titles only, which is insufficient for our work. Therefore, to convert HTML elements in a privacy policy to the new hierarchical structure based on PP-XML, we conducted a five-step process described in Appendix~\ref{appendix:PP2PP-XML}.

\subsubsection{GDPR Concept Annotation}

Using the process described in Section~\ref{subsubsec:constructing_hierarchy}, we processed all 150 privacy policies in our corpus to get 150 PP-XML documents. Then, we annotated the 150 PP-XML documents with relevant GDPR concepts based on our extended GDPR taxonomy introduced in Section~\ref{subsec:GDPR_taxonomy}. The annotation was done for each title and paragraph element in each PP-XML document, with one or more tags each representing a unique GDPR node in our extended GDPR taxonomy.

For the annotation work, we decided to pay Testin (\url{https://www.testin.net/}), a commercial annotation company providing a dedicated services on AI-oriented data annotation, to ensure the quality of the work due to the substantial annotation workload. To further safeguard quality, 
we designed a multi-stepped quality assurance process. In Step 1, one of the authors of this paper, who has good knowledge of GDPR concepts, first trained four annotators of the company, and then the annotators attempted to annotate 20 PP-XML documents as a pilot. In Step 2, the results of the pilot annotation experiments were checked by the author and feedback was given to the four human annotators so they knew how to refine and align their annotations. In Step 3, the four annotators went ahead to annotate all the remaining PP-XML documents so that each document was annotated independently by two different annotators. In Step 4, a random sample of the annotation results of each annotator was checked by another annotator. Finally, the full annotation results were checked by three authors of this paper, who then discussed the results between them to agree on any different annotations. To quantitatively measure the quality of the four annotators' work, we used the inter-rater agreement of Cohen's Kappa ($\kappa$)~\cite{Cohen1960A}, and observed an average $\kappa$ score of 0.75 across all the annotated documents, indicating that the two independent human annotators had a substantial level of agreement for most documents. For disagreements, we observed that most are about the GDPR concepts `PD CATEGORY', `PD ORIGIN', and `PROCESSING PURPOSES'. The reason is that these concepts tend to be declared together in many privacy policies, making it difficult to separate them. At the end of the annotation work, we finally obtained a fully annotated privacy policy corpus GoPPC-150 following PP-XML and our extended GDPR taxonomy.

\section{Comprehensive Performance Evaluation of GDPR Concept Classifiers}
\label{sec:classifiers_evaluation}

To fill the identified research gaps on performance evaluation of GDPR concept classifiers, we conducted a comprehensive performance evaluation of different classifiers considering five different aspects: two different hierarchical structures of classifiers (LPCN and LCN), different input features (multiple sets of traditional features and a new set of features reflecting the hierarchical context), multiple machine learning (ML) models, and two different sample splitting methods (document-level and segment-level). Since the full combinations are very complicated, we first tested different combinations of input features, ML models and hierarchical architectures to select the best ML model for LPCN and LCN, and then derive 12 types of classifiers representing different input features and hierarchical structures. The performances of 12 types of classifiers were then evaluated and compared for different GDPR concepts and different sample splitting methods. In the following, we first give more details of different aspects of our experimental settings in Sections~\ref{subsec:performance_evaluation:corpus}--\ref{subsec:performance_evaluation:sample_splitting}. Then, Section~\ref{subsec:performance_evaluation:classifiers} summarizes the 12 types of classifiers we evaluated, and Section~\ref{subsec:performance_evaluation:results} explains the performance evaluation results of the classifiers.

\subsection{Corpus Used}
\label{subsec:performance_evaluation:corpus}

Due to the consideration of hierarchical classifiers and contextual features reflecting the hierarchical context, we needed a privacy policy corpus including hierarchical labels. Given our new corpus GoPPC-150 is the first and the only corpus meeting such requirements, we used it for all our experiments. GoPPC-150 covers all 96 nodes at three levels of our GDPR taxonomy, so there are maximum 96 concept classifiers we can construct in the LCN setting. However, for some concepts there are not sufficient annotated labels to support meaningful performance evaluation, so we used only 14 level-1 concepts and 21 level-2 concepts, leading to 35 concepts in total.

\subsection{Hierarchical Structures of Classifiers}

As mentioned in Section~\ref{subsec:related_work:HMTC}, privacy policy concept classification can be regard as an HMTC task and both LCPN and LCN classifiers have been explored, although the former dominated the literature. We have not seen any past studies comparing LCPN and LCN classifiers in a single performance evaluation study, so we decided to cover them in our study.

In addition to the hierarchy related to the GDPR concepts, there is another hierarchy we want to highlight. As explained when introducing GoPPC-150, a privacy policy can have a hierarchical structure with multiple semantic segments (i.e., sections) each of which can include one or more paragraphs and potentially one or more sub-segments. Each segment can have a title (i.e., the heading of a section/subsection). Despite the hierarchical structure of privacy policies, past studies on privacy policy concept classification focused mostly on long texts or paragraphs in the privacy policy without considering the hierarchical structure of the document, e.g., no previous work has considered classifying titles of semantic segments in a privacy policy or used the hierarchical information of the document to help inform classification of paragraphs and sub-segments inside a segment. Missing such contextual information may represent a missed opportunity for improving the classification performance for some nodes. We addressed this issue by classifying both titles and paragraphs, and by introducing contextual features capturing potentially useful information from the parent node and sibling nodes. See the next subsection for more information about such hierarchical contextual features.

\subsection{Input Features}
\label{subsec:performance_evaluation:features}

When classifying a paragraph or title in a privacy policy, we considered the following three sub-groups of features as the input of LCN classifiers: (1) the vector embeddings of the paragraph or title, (2) the vector embeddings of the already processed parent and sibling node of the paragraph or title, and (3) a binary vector representing if some pre-defined keywords associated with each GDPR concept appears or not in the current paragraph or title. The first-sub-group of features cover all vector embeddings used in past studies, and the third sub-group were used by Amaral et al.~\cite{Amaral2022}. The second sub-group are new features proposed by us for this work to capture the hierarchical information between the current node and its parent and the contextual information between the current node and its siblings under the same parent.

For the first sub-group of features on vector embeddings, we first tested three widely used traditional vectorization methods with three different dimensionalities (100-D, 200-D and 300-D): TF-IDF~\cite{salton1988term}, GloVe~\cite{pennington2014glove}, and  fastText~\cite{bojanowski2017enriching}. We also tested combinations of three different pairs of the three traditional vectorization methods, with a 100-D sub-vector for each method and a 200-D vector for the combination. The experimental results showed that 300-D GloVe vectors performed the best for binary classifiers and we implemented the methods as the default vectorization for binary classifiers. We then compared GloVe and the more advanced vectorization method based on PrivBERT~\cite{Srinath2020PrivaSeerPrivBERT} as two alternative choices for our more comprehensive experiments.

For the second sub-group of features on capturing hierarchical and contextual information, we used the vector embeddings of already processed parent and sibling nodes of the current node in the same privacy policy. Considering parent nodes tend to be titles while sibling nodes tend to be paragraphs when the current node is a paragraph, we considered that 100-D is enough for the former and 300-D is suitable for the latter. The vector embedding methods considered were also GloVe and PrivBERT. These features help capture the hierarchical information between the current node and its parent and sibling nodes. Our experiments proved the features can indeed improve the performance of classifiers sometimes as shown in Tables~\ref{table:performance_segment_vs_document_levels} and \ref{table:detail_performance}.

For the third sub-groups of features about occurrences of pre-defined keywords, we used a 96-D binary vector (the number of all nodes in our GDPR taxonomy), where each feature indicates if at least one pre-defined keywords associated with a specific GDPR concept appears in the candidate element at least once. For example, a sentence like ``When Google shares your information'' will be labeled as `DATA SHARING.CONDITION' because it contains the keywords `share' and `when'. Because Amaral et al.~\cite{Amaral2022} did not release their pre-defined keywords and our GDPR taxonomy includes more concepts, we decided to define our own keywords list for each of the 96 concepts. The full list of such keywords can be found in the supplementary data file available at \url{https://github.com/tp-sh/GDPR_privacy_policies/blob/main/data/keyword_list.csv}). Our experimental results in Tables~\ref{table:performance_segment_vs_document_levels} and~\ref{table:detail_performance} show that these keyword-based features could help improve the performance of LCN classifiers.

For the LCPN classifiers, we set two sub-groups of input features --  PrivBERT embeddings~\cite{Srinath2020PrivaSeerPrivBERT} of current node's text and PrivBERT embeddings of the parent's and sibling's text's embeddings. The experiments showed the parent's and sibling's embeddings had few help in classifications, so we used the PrivBERT embeddings of current node's text as the input features of the fine-tuned PrivBERT. We did not use keywords-based binary features because LCPN classifiers cover multiple concepts so keywords are less indicative.

It deserves mentioning that because of the short length of most titles (often just a few to several words), using more complicated features are not convenient and sometimes impossible. Our experiments showed that using binary RF models with 100-D TF-IDF vectorization input features are sufficient to support title classifiers with good performance. Therefore, our more comprehensive study focused on paragraph classification only, although the contextual features of titles are used when relevant.

\subsection{ML Models}

For LCN classification, Torre et al.~\cite{torre2020ai} reported that the SVM model achieved the best performance. Therefore, we adopted the SVM model for our testing and conducted comparison experiments using other mainstream ML models, including random forest (RF), XGBoost, ET and neural network (MLP). We used 300-D TF-IDF features of current node as input to test the performance of these models, and the results showed that RF achieved the best performance for the most concept. The full results of all classifiers can be found at \url{https://github.com/tp-sh/GDPR_privacy_policies/blob/main/data/comparison_experiments.csv}. Therefore, we took RF classifiers as the representative of LCN classifiers. For LCPN classification, a multi-class classifier is required to classify all the child nodes under the same parent node, with each class assigned a probability and a prediction made if it exceeds the threshold of 0.5. In the current literature, most studies~\cite{Harkous2018Polisis, Mustapha2020PPC-XLNet, rahat2021automated, Srinath2020PrivaSeerPrivBERT} use multi-class neural network models as the classifier. Therefore, we opted for multi-class neural network as the representative of LCPN classifiers.

\subsection{Sample Splitting Methods}
\label{subsec:performance_evaluation:sample_splitting}

When splitting the corpus into a training set and a testing set, there are two ways to do the splitting: 1) document-level -- splitting the two sets by privacy policies so that all segments/titles/paragraphs in a single privacy policy are either in the training or the testing set, but not both; 2) segment-level -- splitting all segments/titles/paragraphs independent of which privacy policy they belong to so the same privacy policy may have some of its segments/titles/paragraphs in the training set and others in the testing set. We have noticed that most past studies on privacy policy concept classification used the document-level sample splitting method, but some (e.g., \cite{Nejad2020baselinePPC, Mustapha2020PPC-XLNet, Srinath2020PrivaSeerPrivBERT}) used the segment-level splitting method according to the publicly available dataset provided in the official repositories.\footnote{\url{https://github.com/SmartDataAnalytics/Polisis_Benchmark} and \url{https://github.com/euranova/privacy-policy-classification-xlnet}.} 

\begin{figure*}[!htb]
\centering
\includegraphics[width=0.85\linewidth]{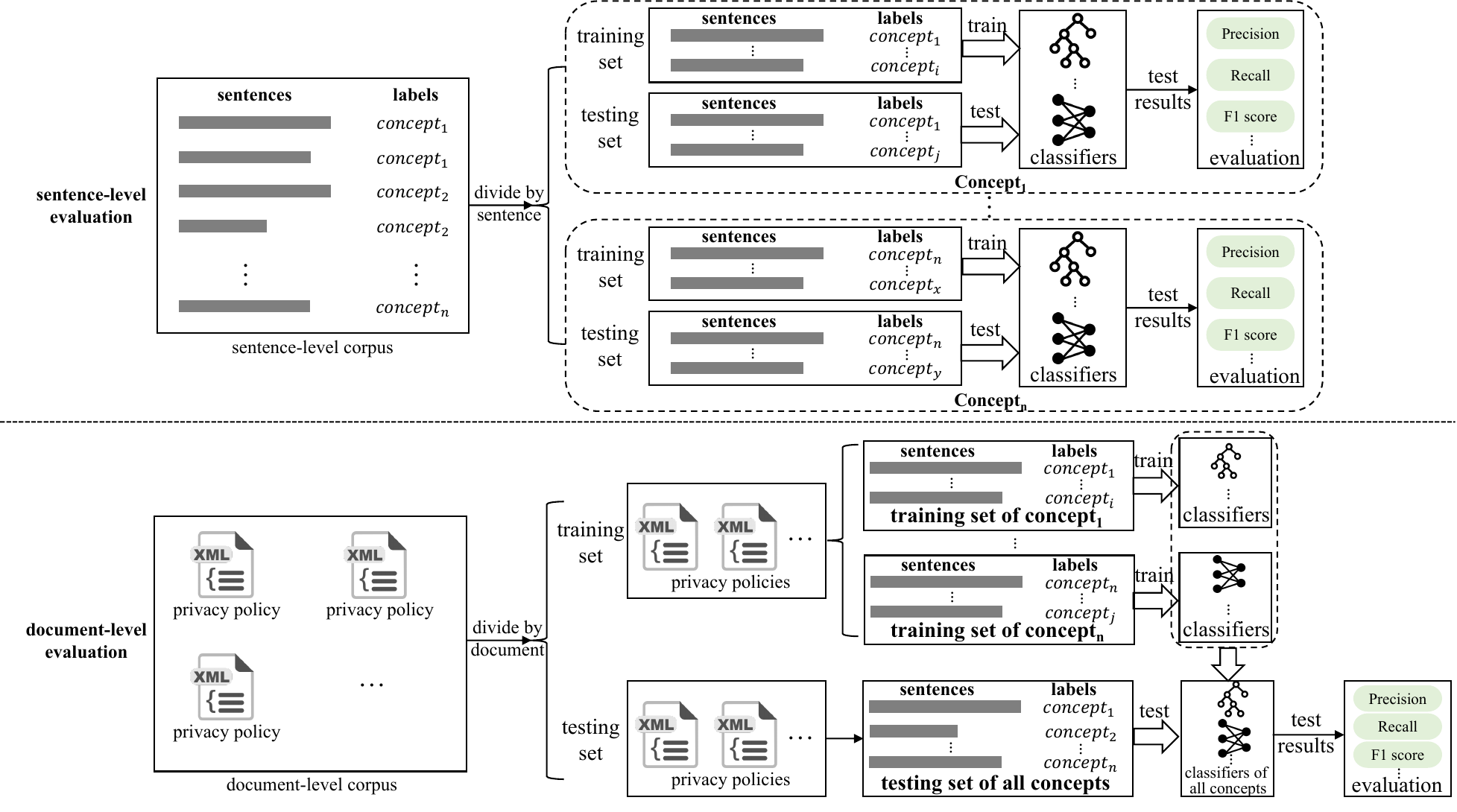}
\caption{Evaluation methods}
\label{fig:sample_splitting_methods}
\end{figure*}

Between the two methods, we believe that the document-level method is more appropriate for the privacy policy concept classification task based on the following fact: in real-world applications, what a trained classifier will not see is usually the whole input privacy policy, not just segments of the privacy policy. Although this belief is supported by most researchers who conducted related studies, the segment-level method remains a possible option for the performance evaluation, and it may be justified by the fact that many privacy policies are written in a similar way (e.g., based on the same template). In the literature, we have not seen any study explicitly investigating the impact of the sample splitting method on the GDPR concept classifiers' performances, so we decided to include them in our study. Figure~\ref{fig:sample_splitting_methods} illustrates how the performance evaluation pipeline differs when the sample splitting method changes.
For all our experiments, We used a 4:1 split between the training and testing sets. 


\subsection{Final Classifiers Evaluated}
\label{subsec:performance_evaluation:classifiers}

By trying different combinations of input features, classifier structures and models, we finally got 12 types of classifiers reflecting different configurations our experiments cover. See Table~\ref{table:6types_classifiers} for more details about the 12 types of classifiers.

\begin{table*}[!htb]
\centering
\small
\setlength{\tabcolsep}{3pt}
\caption{The 12 different types of classifiers}
\label{table:6types_classifiers}
\begin{tabular}{cccc}
\toprule
Type & Input Features & Structure & ML Model\\
\midrule
1 & 300-D GloVe features of the current node & LCN & Binary RF\\
2 & 300-D GloVe features of the current, parent and sibling nodes & LCN & Binary RF\\
3 & 300-D GloVe features of the current node, keyword occurrences & LCN & Binary RF\\
4 & 300-D Glove features of the current, parent and sibling nodes, keyword occurrences & LCN & Binary RF\\
5 & PrivBERT embeddings of the current node & LCN & Binary RF\\
6 & PrivBERT embeddings of the current, parent and sibling nodes & LCN & Binary RF\\
7 & 300-D GloVe features of the current node & LCPN & Multi-class neural network\\
8 & 300-D GloVe features of the current, parent and sibling nodes & LCPN & Multi-class neural network\\
9 & 300-D GloVe features of the current node, keyword occurrences & LCPN & Multi-class neural network\\
10 & 300-D GloVe features of the current, parent and sibling nodes, keyword occurrences & LCPN & Multi-class neural network\\
11 & PrivBERT embeddings of the current node & LCPN & Multi-class neural network\\
12 & PrivBERT embeddings of the current, parent and sibling nodes & LCPN & Multi-class neural network\\
\bottomrule
\end{tabular}
\end{table*}

\subsection{Performance Evaluation Results of Different Classifiers}
\label{subsec:performance_evaluation:results}

We used three common metrics, precision (P), recall (R), and F1-scores, to evaluate the performance of each classifier. The overall performances of the 12 types of classifiers are shown in Table~\ref{table:performance_segment_vs_document_levels} and detailed performances of some examples are shown in Table~\ref{table:detail_performance}. Due to the space limit, we only show the F1-scores of classifiers belonging to 6 representative types in the two tables. The full results of all classifiers can be found at \url{https://github.com/tp-sh/GDPR_privacy_policies/blob/main/data/complete_classifier_results.csv}.

\begin{table*}[!htb]
\centering
\small
\setlength{\tabcolsep}{1pt}
\setlength\extrarowheight{2pt}
\caption{Overall performance results in both segment-level and document-level evaluation frameworks (F1-scores). The rows ``Ratio of maximum F1-score in level-1 nodes'' and ``Ratio of maximum F1-score in all nodes'' indicate the ratio of a specific type of classifiers performed the best among the 14 level-1 nodes and 35 nodes tested in our experiments.}
\label{table:performance_segment_vs_document_levels}
\begin{tabular}{c cccccc | cccccc} 
\hline
\multirow{2}{*}{Evaluation} & \multicolumn{6}{c|}{Segment-level sample splitting} & \multicolumn{6}{c}{Document-level sample splitting}\\
\cline{2-13}
 & Type 1 & Type 2 & Type 3 & Type 4 & Type 11 & Type 12 & Type 1 & Type 2 & Type 3 & Type 4 & Type 11 & Type 12\\
\hline
Macro averages of level-1 nodes & 0.517 & 0.627 & 0.564 & 0.663 & 0.701 & \textbf{0.740} & 0.491 & 0.561 & 0.542 & 0.600 & 0.659 & \textbf{0.669} \\
Macro averages of all nodes & 0.453 & 0.491 & 0.502 & 0.535 & 0.577 & \textbf{0.587} & 0.442 & 0.453 & 0.480 & 0.486 & 0.538 & \textbf{0.542}\\
Ratio of maximum F1-score in level-1 nodes & 0/14 & 1/14 & 0/14 & 1/14 & 2/14 & \textbf{4/14} & 0/14 & 0/14 & 0/14 & 1/14 & \textbf{3/14} & \textbf{3/14}\\
Ratio of maximum F1-score in all nodes & 0/35 & 2/35 & 3/35 & 1/35 & \textbf{8/35} & \textbf{8/35} & 1/35 & 1/35 & 2/35 & 3/35 & 7/35 & \textbf{8/35}\\
\hline
\end{tabular}
\end{table*}

\begin{table*}[!htb]
\centering
\small
\setlength{\tabcolsep}{3pt}
\caption{Detail performance results (F1-scores). For each cell, the performance figure for the document-level evaluation is shown first, followed by that of the segment-level evaluation in parentheses. For LCPN classifiers (Types 11 and 12) on Level 2, their performances are based on those of two cascaded classifiers on Levels 1 and 2. In the ``Name'' column, ``DSR'' represents ``DATA SUBJECT RIGHT'', ``DS'' represents ``DATA SHARING'', and ``LB'' represents ``LAWFUL BASIS''.}
\label{table:detail_performance}
\begin{tabularx}{\linewidth}{cYcccccc} 
\toprule
\multicolumn{2}{c}{Concept} & \multirow{2}{*}{Type 1} & \multirow{2}{*}{Type 2} & \multirow{2}{*}{Type 3} & \multirow{2}{*}{Type 4} & \multirow{2}{*}{Type 11} & \multirow{2}{*}{Type 12}\\
\cmidrule{1-2}
Level & Name & & & & & &\\ 
\midrule
\multirow{5}{*}{1} & CONTROLLER	& 0.544 (0.546)& 0.502 (0.544)& 0.558 (0.623)& 0.558 (0.646)& \textbf{0.584} (0.656)& 0.476 (\textbf{0.705}) \\
& COMPLAINT & 0.609(0.600)& 0.620(0.567)& 0.882(0.825)& \textbf{0.896}(\textbf{0.847})& 0.847(0.828)& 0.826(0.762) \\
& TRANSFER OUTSIDE EUROPE & 0.649 (0.760)& 0.693 (0.770)& 0.662 (0.708)& 0.742 (0.705)& \textbf{0.749} (\textbf{0.871})& 0.716 (0.855) \\
& PD SECURITY & 0.732 (0.627)& 0.768 (\textbf{0.727})& 0.720 (0.588)& 0.781 (0.720)& \textbf{0.803} (0.703)& 0.790 (0.712) \\
& DATA SHARING & 0.599 (0.582)& 0.686 (0.707)& 0.747 (0.707)& 0.754 (0.763)& 0.770 (0.764)& \textbf{0.803} (\textbf{0.810}) \\
\cmidrule{1-8}
\multirow{5}{*}{2} & DSR.ACCESS & 0.482 (0.529)& 0.306 (0.383)& 0.539 (\textbf{0.636})& 0.432 (0.489)& \textbf{0.579} (0.558)& 0.514 (0.459) \\
& DSR.OBJECT & 0.333 (0.480)& 0.279 (0.260)& \textbf{0.442} (\textbf{0.538})& 0.321 (0.380)& 0.404 (0.487)& 0.407 (0.323) \\
& PD ORIGIN.DIRECT ACTIVE & 0.383(0.500)& 0.360(0.498)& 0.412(0.460)& 0.455(0.454)& 0.558(0.514)& \textbf{0.606}(\textbf{0.631}) \\
& LB.CONSENT & 0.211 (0.280)& 0.176 (0.140)& \textbf{0.300} (\textbf{0.291})& 0.188 (0.273)& 0.138 (0.267)& 0.273 (0.286) \\
& DS.RECIPIENT & 0.469 (0.444)& 0.554 (0.538)& 0.542 (0.479)& \textbf{0.580} (0.563)& 0.562 (0.564)& 0.573 (\textbf{0.588}) \\
\bottomrule
\end{tabularx}
\end{table*}

Table~\ref{table:performance_segment_vs_document_levels} shows that Type 12 classifiers (which are PrivBERT-based LCPN classifiers leveraging hierarchical contextual features) performed the best when macro average F1-scores are considered. Such results are not surprising since PrivBERT-based features are the most advanced among all non-contextual features, and the contextual features is expected to be able to help improve classification performance. However, if we look at the results in Table~\ref{table:performance_segment_vs_document_levels}, we can see that, for some GDPR concepts and under different settings, non-PrivBERT classifiers (Types 2--4) could outperform Type 12 classifiers and sometimes Type 11 classifiers also outperformed Type 12 classifiers. Such more nuanced results give a very different picture from what one would expect from Table~\ref{table:performance_segment_vs_document_levels} and results reported in~\cite{Srinath2020PrivaSeerPrivBERT}, which showed that PrivBERT-based classifiers performed always the best for all the OPP-115 concepts, compared with two other types of classifiers.

Comparing the macro average F1-scores of classifiers with the two sample splitting methods shown in Table~\ref{table:performance_segment_vs_document_levels}, we can see that the performance figures with segment-level splitting have a clear tendency to be over-estimated (usually with a significant margin) than with document-level splitting. This is also expected because segment-level splitting will allow the trained classifiers to see more privacy policies fully or partly, therefore giving more opportunities to learn about content of such policies. Looking into F1-scores of specific concept classifiers in Table~\ref{table:performance_segment_vs_document_levels}, this trend largely remains, although there are also exceptions, e.g., for DSR.OBJECT the F1-scores with segment-level splitting are always smaller than those with document-level splitting, across all 6 types of classifiers.

Comparing the performance figures of level-1 and level-2 concept classifiers, we can see a clear pattern that level-2 concept classifiers performed generally much worse than level-1 classifiers. In general, the performances of level-2 classifiers are so poor (often fall below 0.5) that they are not really usable. These results are also not surprising because level-2 concept classifiers are cascaded so there are always cascading errors. In addition, level-2 concepts generally have less data available in the corpus, so the classifiers are likely not sufficiently trained. How to improve them will be a major direction for future research. 

Among all the results, the failure of the contextual features for some concepts were the least expected. This phenomenon may be explained as follows: some concepts are ambiguous and contextual features can help distinguish them from others, but for some other concepts contextual features may just add more redundant or misleading information and confuse the learning process. Further refinement of the contextual features may help make them work better. More future research will be needed to gain more insights.

\section{Further Discussions}
\label{sec:further_discussions}

Our experimental results showed that the actual performances of GDPR concept classifiers have been largely overestimated: although many past studies showed performance over 0.8 or even 0.9 for many concept classifiers, when they are applied to our hierarchical and more complicated corpus (GoPPC-150) and when the document-level sample splitting method is used, the performance can hardly exceed 0.8 even for level-1 concepts and the performance of level-2 concepts is mostly worse than random guesses, making them not practically useful for real-world applications. We believe the performance drop is more about the corpus we used is much more complicated than the simpler ones (e.g., OPP-115 with just 12 concepts), indicating the need for testing all future classifiers with GoPPC-150 to verify their cross-corpus performances. If the performance drop is indeed true, then  simply applying such classifiers to large-scale analyses of privacy policies can lead to too many false positives and false negatives, therefore their conclusions need to be understood with great caution.

Another major insight learned from our experimental results is that there is no a single ``winning'' type of classifiers that always perform the best for all concepts. For instance, as shown in Table~\ref{table:detail_performance}, although Type 12 classifiers largely dominate, for some concepts other types of classifiers performed the best, e.g., for COMPLAINT Type 4 classifiers (LCN classifiers using GloVe features and hierarchical contextual features) outperformed Type 12 ones with a significant margin. Such results may indicates that a ``one size fits all'' approach does not work so more concept-specific classifiers are needed. They may also be an experimental artifact of insufficient training data because Type 12 classifiers may simply need more training data to perform better for some concepts, which our corpus GoPPC-150 cannot provide with just 150 privacy policies.

The significant differences of the performance results obtained by the two sample splitting methods demonstrates that using the segment-level method may lead to an over-estimation of the performance in most cases. Since the document-level evaluation method matches real-world applications better, it should be consistently used for conducting future research on privacy policy concept classifiers.

Our work focused on GDPR-related concepts. Although other countries' data protection laws share many concepts with the GDPR, our GDPR-oriented taxonomy cannot be directly used to cover other data protection laws.

As a whole, our results call for more research on the following important areas: 1) further extending the GDPR taxonomy we produced to cover more data protection laws and concepts; 2) constructing large corpora to produce more data for training privacy policy concept classifiers: and 3) exploring more advanced techniques to further improve the performance of privacy policy concept classifiers, especially those requiring less or even no training data. For the second area, we call the community to help further extend GoPPC-150 to include more privacy policies with manually verified annotations. Our framework for constructing GoPPC-150 and the source code we released can help facilitate such a community-wide effort. For all the areas, the use of large language models (LLMs) is a natural choice since they have been proven capable of conducting many natural language tasks of similar nature under zero/few-shot settings. Possible ideas include using LLMs to analyze data protection laws from different countries and to automatically suggest changes to our GDPR concept taxonomy, using LLMs to automate the construction of a larger corpus, and using LLMs to develop classifiers with few-short training setting, via fine-tuning or facilitated by RAG (retrieval-augmented generation). Since 2024, some researchers have started exploring the use of LLMs for automating analysis of privacy policies~\cite{Rodriguez2024LLMPPAnalysis, Yang2025PPAnalysisLLM, Xie2025LLMPrivacy}, but they are not about the above areas we consider more fundamentally important, e.g., they still used smaller privacy policy corpora and did not consider the hierarchical information in privacy policies.

\section{Conclusion}
\label{sec:conclusion}

This paper reports our work on an extended GDPR-oriented concept taxonomy, a new privacy policy corpus with hierarchical information from 150 privacy policies, and results from a comprehensive performance evaluation study of GDPR concept classifiers considering multiple aspects of such performance evaluation experiments. The taxonomy supported the development of the corpus, as part of a framework for constructing such corpora, and the new corpus made it possible for us to consider two hierarchical structures and some hierarchical contextual features for the GDPR classifiers. Multiple traditional features and machine learning models were also considered, together with two sample splitting methods. Our comprehensive performance evaluation study led to a range of new findings and insights about performances of machine learning based GDPR concept classifiers, including the usefulness of different hierarchical structures and hierarchical contextual features, the observation that some simpler classifiers may outperform more advanced classifiers, and the finding that a ``one size fits all'' idea may not work for privacy policy concept classifiers. Our study led to a range of new GDPR concept classifiers that can be used by researchers and practitioners as better baselines. Although our work focused on GDPR-oriented analysis, most techniques, tools and processes developed can either be easily extended to study other and multiple data protection laws.

\appendices

\section{Full list of HTML elements removed}
\label{appendix:HTML_elements_removed}

The following HTML elements are removed when pre-processing the web page:

\begin{itemize}
\item \emph{Type 1}: Elements that are not useful for analyzing textual content of the privacy policy contained in the web page. These include \verb|<img>|, \verb|<picture>|, \verb|<video>|, \verb|<audio>|, \verb|<canvas>|, \verb|<map>|, \verb|<area>| \verb|<figure>|, \verb|<figcaption>|, \verb|<source>|, \verb|<track>| and \verb|<svg>| elements.

\item \emph{Type 2}: Elements that embed an non-textual object or an external web page. These include \verb|<applet>|, \verb|<embed>|, \verb|<object>|, \verb|<param>|, \verb|<script>|, \verb|<noscript>| and \verb|<iframe>| elements.

\item \emph{Type 3}: Elements whose are not semantically related to or can appear as part of the main body of a web page. These include \verb|<footer>|, \verb|<nav>|, \verb|<form>| and all input control elements.
\end{itemize}

\section{The process of constructing hierarchical structure}
\label{appendix:PP2PP-XML}

\emph{Step 1: Pre-processing.} Although many HTML elements have been removed as part of the first pre-processing step described in Section~\ref{subsubsec:pre-processing-PP-web-page}, some ad hoc inline elements, such as those appearing in the middle of a \verb|<p>| element, play no roles in PP-XML. We remove such inline elements. At the end, we put all remaining elements into a \verb|<policy>| element.

\emph{Step 2: Initial conversion of some HTML elements to PP-XML elements.} As shown in Table~\ref{tab:PP-XML-elements}, there are a precise mapping from some HTML elements to \verb|<list>| and \verb|<item>| elements in PP-XML, so such HTML elements can be directly converted.

\begin{table*}[!t]
\centering
\small
\caption{Different types of PP-XML elements and HTML elements that may be converted to each PP-XML element type}
\label{tab:PP-XML-elements}
\begin{tabularx}{\linewidth}{rXp{0.3\linewidth}}
\toprule
PP-XML element & Description & Possible HTML elements\\
\midrule
\verb|<policy>| & The privacy policy as a whole (the root element) & NA (determined based on the method introduced in Section~\ref{subsubsec:extracting_PP_content})\\
\verb|<segment>| & A semantic segment of the privacy policy (child of the \verb|<policy>| element or another \verb|<segment>| element) & NA (derived from \verb|<title>| elements, see Step~4 discussed in Section~\ref{subsubsec:constructing_hierarchy})\\
\verb|<title>| & The title of a semantic segment (child of a \verb|<segment>| element) & \verb|<h1>|, \verb|<h2>|, \verb|<h3>|, \verb|<h4>|, \verb|<h5>|, \verb|<h6>|, \verb|<p>|, \verb|<div>|, and highlighted inline elements\\
\verb|<paragraph>| & A semantic paragraph in the privacy policy, normally but not always in a semantic segment (child of a \verb|<segment>| element or the \verb|<policy>| element) & Block-level elements such as \verb|<p>| and \verb|<div>|\\
\verb|<list>| & A semantic list in the privacy policy (child of a \verb|<paragraph>| element or an \verb|<item>| element) & \verb|<ul>|, \verb|<ol>|, \verb|<dd>|\\
\verb|<item>| & An item in a semantic list (child of \verb|<list>| element) & \verb|<li>|, \verb|<dt>|\\
\bottomrule
\end{tabularx}
\end{table*}

\emph{Step 3: Classifying the remaining HTML elements into title elements at different levels and paragraph elements in PP-XML.} In privacy policies, titles tend to have different attributes from paragraphs, e.g., titles are often in bold face and have a shorter text length. Based on manual inspection of title and paragraph elements of some randomly selected privacy policies, we identified the following attributes that may be useful for differentiating title elements from paragraph elements: i) the text length, ii) the font size, iii) the font weight, iv) if the text is italic, v) if the text is underlined, vi) the node depth in the HTML DOM tree, vii) the HTML tag, and viii) leading ordinal labels (LOLs) for titles (e.g., `1', `2.3', `a.', and `ii)') (see Table~\ref{table:serial_number_of_labels'_format} for more examples on LOLs). As part of the input features of the title and paragraph classifier, a 12-D descriptor is used to describe the leading ordinal label of a candidate title or paragraph element. The 12-D vector is composed of four 3-D sub-vectors, each representing a sub-label of the leading ordinal label at one of the four possible title levels. Each sub-vector follows the format (sub-label's format, sub-label's value, sub-separator's format). The sub-label's format and the sub-separator's format are determined according to Table~\ref{table:serial_number_of_labels'_format}. The sub-label's value is the 1-based natural ordinal value of the label, e.g., the number itself for Arabic numbers, 1 for `a', `A' and `i'. For instance, the sub-label `3.' will be mapped to (1, 3, 1), `b)' to (3, 2, 3), and a full label `3.a.i' will be mapped to the 12-D descriptor [1 3 1 2 1 1 4 1 0 0 0 0]. 

In total, for each candidate title/paragraph element, we have these input features and an ML-based classifier can be constructed to predict the candidate element into one of the following five classes: title at Level~$i$ ($i=1,2,3,4$), and paragraphs. We tested several mainstream ML models that do not require a large training set for the multi-class classification task, including random forest (RF)~\cite{WILLIAMS20201639}, SVM (with linear kernel and RBF kernel)~\cite{moguerza2006support}, Extra Trees (ET)~\cite{Geurts2005ExtraTrees}, and XGBoost~\cite{XGBoost}. We used all the 150 privacy policies for training and testing purposes. We used 5-fold cross-validation and 20\% as the testing set. Table~\ref{table:results_title_paragraph_classifiers} shows the results of all classifiers, indicating that the ET classifier achieved the best performance, with the F1-score reaching 0.882. As can be seen, the performance is generally good enough to support the semi-automated corpus construction process.

\emph{Step 4: Constructing segment elements based on title elements and their levels.} After the \verb|<title>| elements and their levels are identified, we can use them to construct properly nested \verb|<segment>| elements reflecting the hierarchical structure of the privacy policy. For each \verb|<title>| element, we create a \verb|<segment>| element to include the \verb|<title>| element and all non-\verb|<title>| elements after it until the next \verb|<title>| element or the end of the PP-XML document. The level of the \verb|<segment>| element is set to be the same as the \verb|<title>| element it contains.

\emph{Step 5: Manual verification and correction.} The last step of the process is to have one or more human experts to manually check the produced PP-XML document and the input HTML document to confirm the automated results and fix any errors.

\begin{table}[H]
\centering
\small
\caption{Mappings between sub-label's format and sub-separator's format and values in the leading ordinal label descriptor}
\label{table:serial_number_of_labels'_format}
\begin{tabular}{ccc}
\toprule
Descriptor Value & Label's Format & Separator's Format\\
\midrule
0 & None & None\\
1 & Arabic Number & Full Stop\\
2 & Lowercase Letter & Colon\\
3 & Uppercase Letter & Parenthesis\\
4 & Roman Number & Others\\
5 & Others & Others\\
\bottomrule
\end{tabular}
\end{table}

\begin{table}[H]
\centering
\caption{Results of title and paragraph classifiers}
\label{table:results_title_paragraph_classifiers}
\begin{tabularx}{\linewidth}{ccYYcc}
\toprule
Classifier & RF & SVM (Linear) & SVM (RBF) & XGBoost & ET\\
\midrule
Precision & 0.874 & 0.879 & 0.803 & 0.711 & \textbf{0.885}\\
Recall & 0.875 & 0.873 & 0.822 & 0.778 & \textbf{0.879}\\
F1-score & 0.875 & 0.876 & 0.812 & 0.742 & \textbf{0.882}\\
\bottomrule
\end{tabularx}
\end{table}

\section{Complete result}

Table \ref{table:result_all_types} shows the results of 12 type classifiers in both segment and document level evaluation.

\begin{table*}[!htb]
\centering
\small
\setlength{\tabcolsep}{6pt}
\setlength\extrarowheight{2pt}
\caption{F1-scores under segment-level and document-level evaluation frameworks (all model types), reported as Macro and Micro averages.}
\label{table:result_all_types}
\begin{tabular}{lcccc|cccc}
\hline
\multirow{3}{*}{Type} &
\multicolumn{4}{c|}{Segment-level} &
\multicolumn{4}{c}{Document-level} \\
\cline{2-9}
& \multicolumn{2}{c}{Level-1 nodes} & \multicolumn{2}{c|}{All nodes} &
\multicolumn{2}{c}{Level-1 nodes} & \multicolumn{2}{c}{All nodes} \\
\cline{2-9}
& Macro & Micro & Macro & Micro & Macro & Micro & Macro & Micro \\
\hline
Type 1  & 0.517 & 0.468 & 0.453 & 0.460 & 0.491 & 0.452 & 0.442 & 0.446 \\
Type 2  & 0.627 & 0.606 & 0.491 & 0.545 & 0.561 & 0.545 & 0.453 & 0.501 \\
Type 3  & 0.564 & 0.523 & 0.502 & 0.506 & 0.542 & 0.499 & 0.480 & 0.489 \\
Type 4  & 0.663 & 0.632 & 0.535 & 0.577 & 0.600 & 0.574 & 0.486 & 0.534 \\
Type 5  & 0.697 & 0.689 & 0.583 & 0.616 & 0.657 & 0.659 & 0.529 & 0.598 \\
Type 6  & 0.736 & 0.722 & \textbf{0.604} & 0.643 & \textbf{0.685} & 0.676 & \textbf{0.553} & 0.603 \\
Type 7  & 0.586 & 0.588 & 0.412 & 0.509 & 0.558 & 0.564 & 0.418 & 0.503 \\
Type 8  & 0.670 & 0.674 & 0.485 & 0.590 & 0.637 & 0.656 & 0.456 & 0.575 \\
Type 9  & 0.621 & 0.613 & 0.494 & 0.552 & 0.601 & 0.584 & 0.472 & 0.541 \\
Type 10  & 0.693 & 0.694 & 0.531 & 0.622 & 0.665 & 0.658 & 0.525 & 0.601 \\
Type 11  & 0.701 & 0.685 & 0.577 & 0.637 & 0.659 & 0.666 & 0.538 & 0.606 \\
Type 12  & \textbf{0.740} & \textbf{0.753} & 0.587 & \textbf{0.684} & 0.669 & \textbf{0.686} & 0.542 & \textbf{0.612} \\
\hline
\end{tabular}
\end{table*}

\bibliographystyle{IEEEtran}
\bibliography{main}

\end{document}